\documentstyle[12pt,epsf]{article}
\def\scsc{\scriptscriptstyle}
\def\goes{\rightarrow}
 
      \def\mz{M_{Z}}
      \def\mw{M_{W}}

                 \def\tb{\tan \beta}

                 \def\msll{m_{{\widetilde \ell}_L}}
                 \def\mslr{m_{{\widetilde \ell}_R}}
                 
                 \def\mser{m_{{\widetilde e}_R}}
                 \def\msnu{m_{\widetilde \nu}}
                 \def\msl{m_{\widetilde \ell}}
                 \def\at{A_{t}}
                 \def\ab{A_{b}}
      
      \def\sq{\widetilde q}
      \def\stl{{\widetilde t}_L}
      \def\str{{\widetilde t}_R}
      \def\sbl{{\widetilde b}_L}
      \def\sbr{{\widetilde b}_R}
      \def\slel{{\widetilde e}_L}
      \def\sler{{\widetilde e}_R}
      \def\slep{\widetilde \ell}
      \def\sll{{\widetilde \ell}_L}
      \def\slr{{\widetilde \ell}_R}
      \def\snu{\widetilde \nu}

\begin{document}
                 \setcounter{page}{0}
                 \thispagestyle{empty}
\begin{flushright} TIFR/TH/96-05 \end{flushright}

\begin{center} 
{\Large\bf 

Higgs-mediated Slepton Pair-production at the Large Hadron Collider.}
\\

\vspace{0.5in} \normalsize

Mike Bisset$^a$, Sreerup Raychaudhuri$^a$ and Probir Roy$^a$ \\ 
{\it Theoretical Physics Group, Tata Institute of Fundamental
Research, \\ Homi Bhabha Road, Bombay 400 005, India.} \\
\vskip 30pt

{\Large\bf Abstract}

\end{center}

At the LHC, directly pair-produced sleptons may be easier to identify
than those arising in cascade decays of squarks or gluinos. Higgs
exchange processes leading to such slepton pair-production are
calculated to one-loop in the MSSM. It turns out, surprisingly, that 
their total contribution can dominate the usual Drell-Yan production in
certain regions of the parameter space.
In particular, an interesting region with low $\tan
\beta$ is covered by the dominant exchange of the heavier neutral
scalar $H^0$.

\bigskip

\vskip 5cm

{\footnotesize 
\noindent $^a$email: 
~bisset@theory.tifr.res.in; ~ ~sreerup@theory.tifr.res.in; ~
~probir@theory.tifr.res.in }

\newpage
The Minimal Supersymmetric Standard Model (MSSM) \cite{SUSYreviews}
is a promising candidate for physics beyond the Standard Model. If
supersymmetry is to stabilize the electroweak scale and the masses of
fundamental scalar fields, then the sparticles should have masses
below a few TeV \cite{sparticlemasses}. This is also the mass window
in which the Higgs bosons of the MSSM are expected to lie.  It is
expected, therefore, that supersymmetry could be discovered at the
Large Hadron Collider (LHC).  The lightest sparticle (LSP), presumed
to be the lowest mass neutralino and stable on account of $R$-parity
conservation, would escape detection leading to distinctive missing
transverse energy $({E\!\!\!\!/}_T)$ signals.  The absence of such 
signatures has
already enabled the lower energy LEP-1, LEP-1.5 and Tevatron colliders
to generate
constraints on sparticle masses and couplings.  In this letter 
one clear signal for supersymmetry is discussed, namely, direct
electroweak hadroproduction of a pair of charged or neutral sleptons which
subsequently decay to acollinear unlike-sign 
but like-flavor dileptons with ${E\!\!\!\!/}_T$.
Earlier studies with respect to the Tevatron \cite{sleptonsignals}
have shown a large Standard Model background from $W^+W^-$ production
swamping this signal. At the LHC, however, the event rate is
expected to be larger (because of the higher energy and gluon
luminosities) so that more stringent kinematic cuts can be applied to
eliminate backgrounds leaving \cite{sleptonsignals} a visible and
tractable signal.

Sleptons might also appear in cascade decays of squarks and
gluinos. However, if the latter are close to $1 \, \hbox{TeV}$ in mass, they
would be beyond the reach of the Tevatron --- even after a
luminosity upgrade --- and would be produced in small numbers in the
initial runs of the LHC due to phase-space limitations. On
the other hand, sleptons can be as light as a hundred GeV or so.  (In
supergravity-constrained  models, for example, light slepton masses are 
expected
if the squark and gluino masses are roughly equal.) Searching for
directly produced low-mass slepton pairs may, then, be more fruitful
than waiting for enough squark and gluino events to build up for
cascade decay signals to be viable.  Alternatively, squarks and
gluinos could be of lower mass and hence be produced bountifully in
hadron colliders. For instance, with masses 
$\sim$ $200 \, \hbox{GeV}$, they
could be seen at the upgraded Tevatron. Even heavier 
($\sim$ $500 \, \hbox{GeV}$)
ones could be produced copiously at the LHC. Sleptons could then
occur in cascades, but would not be easily distinguishable
\cite{sleptonsignals} from cascade-produced charginos or neutralinos.
In contrast, such a distinction seems to be achievable
\cite{sleptonsignals} in the case of direct electroweak production of
slepton pairs. Moreover squarks and gluinos may also decay directly 
into
LSP's with very small branching fractions for sleptonic modes. There
are two additional advantages for directly produced slepton pairs:
($i$) with only hard multileptons detected, the events would be
hadronically quiet, so that cuts on hadronic activity could 
suppress backgrounds such as $t\bar{t}$, provided there is an efficient veto
on soft jets; ($ii$) leptons arising
from directly pair-produced sleptons must be of the same flavor
unlike those coming from $W^+W^-$. Considering all these aspects, it
is clearly worthwhile to look at direct electroweak pair-production
of sleptons at hadron colliders. A complete and reliable 
calculation of the total electroweak production rate is first required 
before the signals for sleptons can be analyzed.  

Any of the following four mechanisms can lead to significant 
slepton pair-production:
($i$) gluon fusion through a top or stop loop to produce a 
$s$-channel neutral Higgs boson intermediary, $h^0$, $H^0$, or $A^0$, 
which goes into a slepton pair\footnote{The amplitude for $Z^0$-mediated 
gluon fusion into a slepton pair vanishes \cite{oldgluonfusion}. 
Also, $A^0$-exchange contributes only to the $\sll^\pm \slr^\mp$, 
amplitudes which are suppressed by $m_{\ell}/\mw$ in the coupling.};
($ii$) Drell-Yan production in which a $q\bar{q}$
pair goes into a slepton pair through $s$-channel
$\gamma$ and/or $Z^0$ exchange or a pair of dissimilar quarks $q
\bar{q}^\prime$ form a charged-neutral slepton pair ($\slep^-
\snu_{\ell}^*$, $\slep^+ \snu_{\ell}$) through $W^\pm$-exchange; 
($iii$) $W^+W^-$ fusion; ($iv$) $b\bar{b}$ (or $c\bar{c}$) sea-quark
annihilation to slepton pairs through $s$-channel exchange of neutral scalars
$h^0,H^0,A^0$.  Though ($iv$) has not been discussed in the literature, 
($i$), ($ii$), ($iii$) have been \cite{EHLQ,sleptonsignals,oldgluonfusion}.
In particular, del Aguila and Amettler \cite{oldgluonfusion} found the
Drell-Yan mode to be dominant, $W^+W^-$ fusion to be an
order of magnitude smaller (in general) 
and gluon fusion to be smaller still.  In
their analysis, only the exchange of the lightest neutral Higgs, $h^0$,
was considered in gluon fusion; the two other neutral Higgs bosons $H^0,A^0$
were assumed to be very heavy ($\sim$ $1 \, \hbox{TeV}$) and neglected.  
However,
large regions of the MSSM parameter space exist where $H^0$ is much
lighter and can make a substantial contribution to the gluon fusion
mode of production for sleptons.  The analysis by the authors of Ref.
\cite{oldgluonfusion} was, therefore, incomplete.  Moreover, they
did not consider 1-loop radiative corrections to Higgs
masses which are now known to push the upper bound on $m_h$
to about $130 \, \hbox{GeV}$ \cite{topstoploops}; this, in fact, 
increases the $h^0$-contribution to the process.  Subsequently Ref.
\cite{sleptonsignals} concentrated on Drell-Yan production
only, taking the conclusions of Ref. \cite{oldgluonfusion} to
justify their neglect of gluon fusion.

The results of a more complete analysis of slepton
pair-production are presented here.  
The $h^0$-contribution --- even after 1-loop corrections,
remains small, exceeding the Drell-Yan part only for slepton
masses between roughly $50$ and $60 \, \hbox{GeV}$. This can be seen in the 
high $\tan\beta$ curves in Fig. 1($a$). In contrast, $H^0$-exchange which 
covers an interesting part of the parameter space with low $\tan \beta$ and
$m_A$ (consistent with supergravity constraints \cite{cMSSM}) turns
out to be dominant and yields a cross-section which is significantly
larger than that from all other mechanisms including the Drell-Yan
one. Thus slepton pairs can be produced more copiously
than was thought previously; moreover, the kinematic distribution of the
produced sleptons may differ from the Drell-Yan case which
might prove an aid to detection.

Gluon fusion to Higgs bosons takes place through quark $(q)$ and
squark ($\sq$) mediated one-loop diagrams \cite{Guide}.  Since the
Yukawa couplings of the Higgs bosons to $q(\sq)$ pairs are
proportional to $m_q/\mw$, the $t(\stl,\str)$ loops dominate though
all $q ({\sq})$ loops have been included in the numerical 
calculation. Off-shell effects from the exchanged Higgs bosons have
been included; however,
the dominant contributions come when the Higgs bosons are on-shell, since the
resonances are rather narrow. Interference terms
involving different scalars in the propagator are, therefore, small
and have not been presented in the formulae below.

The Higgs interactions relevant here are
\begin{equation}
{\cal L}^{Higgs}_I   = \sum_{S = h,H} e \Big[ {\cal
A}_{S{\slep}\slep^*} S {\slep} {\slep}^* + {\cal A}_{S{\sq}\sq^*} S
{\sq} {\sq}^* + S \bar{q}  ({\cal A}_{S{q}\bar{q}} + {\cal
A}^\prime_{S{q}\bar{q}} \gamma_5) q  \Big] + {\rm h.c.} \; ,
\end{equation}
\noindent where the couplings 
${\cal A}_{S{\slep}{\slep}^*}, {\cal A}_{S{\sq}{\sq}^*}, {\cal
A}_{S{q}\bar{q}}, {\cal A}^\prime_{S{q}\bar{q}}$ can be easily read
off from Ref. \cite{MikeThesis}.  We use the symbol $\slep$
generically for both sneutrinos $\snu$ and charged sleptons
$\slep^\pm_L, \slep^\pm_R$.  The slepton pair-production
cross-section through gluon fusion is
\begin{eqnarray}
& & \sigma_{gluon} (pp \goes \slep {\slep}^* + X) 
 =  \frac{1}{16 \pi s^2} \sum_{S = h,H} 
\int \frac{dx_1}{x_1^2} \frac{dx_2}{x_2^2} ~f_{p/g}(x_1) ~f_{p/g}(x_2)
\int d\hat{t} ~\mid \overline{\cal M} \mid^2_{gluon},
\\
& & \mid\overline{\cal M}\mid^2_{gluon} 
 =  \frac{1}{4} \frac{\alpha \alpha_s^2 
\mid {\cal A}_{S{\slep}\slep^*} \mid^2 } 
{(\hat{s} - m_S^2)^2 + (m_S \Gamma_S)^2}
\nonumber \\
& & ~ ~ ~ ~ ~ ~ ~ ~ ~ ~ \times \Big[ \mid 
  \tau_q^{-1} {\cal A}_{S{q}\bar{q}} F_{1/2}(\tau_q) 
+ (2\tau_q)^{-1} {\cal A}_{S{\sq}\sq^*}  F_0 (\tau_{\sq}) \mid^2
+ ~ m^2_S \tau_q \mid {\cal A}^\prime_{S{q}\bar{q}} f(\tau_q) \mid^2
           \Big],
\nonumber
\end{eqnarray}
where $\tau_q = 4 m_q^2 / \hat{s}$, $\tau_{\sq} = 4 m_{\sq}^2 /
\hat{s}$ and the functions $F_{1/2}, F_0, f$ are given in Appendix C
of Ref. \cite{Guide}. In the above, $s$ is the square of the $pp$
center-of-mass energy and $\hat{s},\hat{t}$ are the Mandelstam
variables for the subprocess in the parton center-of-mass frame.
Notations for the mass parameters are self-explanatory.

While calculating the widths
\cite{energydependentwidths} of the scalars $S = h^0,H^0$ in the
Breit-Wigner form of the propagator, all possible
decay channels allowed by the respective masses for the Higgs bosons
are included.  The masses are calculated in terms of the MSSM parameters
including radiative corrections to the self-energies from 
$t ~(\stl,\str), b ~(\sbl,\sbr)$ loops\footnote{These corrections are
relatively small for the $H^0$ which makes the largest contribution to
gluon fusion in the interesting region of parameter space.}. 
Parton distributions are taken from the recent MRS(A$^\prime$)
parametrization of Martin, Roberts and Stirling
\cite{structurefunctions}, though their numerical values do not
differ significantly from earlier parametrisations ({\it e.g.} those
used in Ref. \cite{oldgluonfusion}) at the scale $Q^2 \sim 10^4 \,
\hbox{GeV}^2$. Both left and right charged sleptons are considered.
Formulae similar to (2) hold for the cases of Drell-Yan production
and $b\bar{b}$ annihilation; these will be presented elsewhere.
Finally, $W^+W^-$ fusion terms, calculated using the
effective-$W$ approximation in the limit of high energies
\cite{effectiveW}, are at least an order of magnitude smaller than
Drell-Yan production and can be safely ignored.

QCD corrections to the gluon fusion amplitude have not been included.
This makes our estimates of Higgs exchange rates somewhat
conservative since these corrections are known to {\it increase} gluon
fusion rates by a factor as large as 1.5 \cite{QCDcorrections}. 
QCD corrections to Drell-Yan production of slepton pairs, which result 
mainly from the annihilation of light quarks, have been neglected;
those to lepton pair-production at these energies are known 
\cite{DrellYanQCDcorrections} to be small and these are expected to
be reduced further for heavier mass sleptons.
However, for $b \bar{b}$ annihilation, radiative corrections 
\cite{radiativecorrections} drive the cross-section {\em down} to about 
a quarter of the tree-level result; these corrections have, therefore, 
been incorporated in our calculation.

The dominance of Higgs exchange over the Drell-Yan process as a
production mode in a region of the allowed parameter space is
illustrated in Figs. 1 ($a$-$c$). In Fig. 1($a$) cross-sections for 
the production of {\it one} species of sneutrino are plotted against 
its mass $\msnu$. The mass $m_A$ of the pseudoscalar 
Higgs has been fixed at $225 \, \hbox{GeV}$ for definiteness;
other input parameters are listed in the next paragraph.  Solid lines
correspond to the total Higgs-mediated cross-sections (gluon fusion
plus $b \bar{b}$ annihilation) for the marked values of $\tb$.
The values $1$ and $1.5$ of $\tb$ may not be completely consistent with the 
full supergravity constraints at the grand unification scale and
radiative electroweak symmetry-breaking, but are allowed if 
the universal scalar mass assumption is partially relaxed.
Though gluon fusion with $H^0$ exchange is the dominant mechanism, the
$b \bar{b}$ annihilation process gives about a third as much for $\tb
\simeq 3 - 5$. The dashed line shows the Drell-Yan cross-section
with $\gamma$ and $Z^0$ exchanges, which is independent\footnote{This
is strictly true for sneutrinos where there is no left-right mixing.
For charged selectrons there is a dependence on $\tb$ which appears
in the mixing matrix, but this is very weak since the relevant term
is suppressed by $m_{\ell}/\mw$.} of $\tb$. Evidently, given a low $\tb$, 
Higgs
exchange is dominant for $\msnu$ from $60-70\, \hbox{GeV}$ to around
$120\, \hbox{GeV}$ at which point the exchanged $H^0$ goes off-shell.
With increasing $\tb$, this cross-section falls quite dramatically
--- to a negligible level for $\tb > 3$. Thus $b \bar{b}$
annihilation, which is significant only in this region, {\em is
always a small part of} the net cross-section (see Table 1).  In Fig.
1($b$) a similar plot for the pair-production of right selectrons has 
been shown using the same conventions. Gluon fusion dominates the
Drell-Yan mode for $\mser > 80\, \hbox{GeV}$ and $\tb \simeq 1$ and
remains comparable upto $\tb = 2$.  Fig. 1($c$) illustrates the
situation for left selectron pairs, where the two modes are
comparable at best and Drell-Yan production becomes dominant for $\tb
> 1$.

We now comment on the choice of parameters. Heavy quark masses are 
taken to be $m_t (m_b) = 175 (5) \, \hbox{GeV}$; all squark masses are 
set to $1 \, \hbox{TeV}$ and the gluino mass is $900 \, \hbox{GeV}$, 
while $\mu = -250$ GeV and $\at = \ab = -500 \, \hbox{GeV}$. 
This constitutes a reasonable choice of parameters in the MSSM, but
is not the optimal one for slepton pair-production.
$SU(2)_L$-invariance predicts the equality of the soft 
supersymmetry-breaking masses of $\sll$ and $\snu_{\ell}$ for each 
generation, leaving six free input parameters in the slepton mass 
sector\footnote{The physical masses, always denoted by $\msl$ in this letter,
become different because of
electroweak symmetry-breaking and mixing, the latter being
significant only for staus. Inputs such as $\tb$ and $\mu$ entering
via these effects are not counted among the six parameters; further,
$A_{\ell}$-terms are neglected.}.  For simplicity, the soft SUSY-breaking
slepton masses are chosen to be flavor-degenerate in the left and right sectors
separately, the latter being slightly greater than the former. 
A typical (not necessarily optimal) soft mass-squared difference of 
$20 \, \hbox{GeV}^2$
between left and right sleptons is taken, corresponding to a
mass difference of about $0.1\, \hbox{GeV}$ for soft masses of 
$\sim$ $100 \, \hbox{GeV}$.  

Since soft-supersymmetry breaking stau masses have been taken to be
degenerate with selectron and smuon soft masses, intra-flavor
left-right mixing drives the lighter physical stau mass below the
other charged slepton masses.  This increases the total Higgs width
in the Breit-Wigner denominator, decreasing the selectron
pair-production rates. If, on the other hand, staus and/or
smuons are much heavier than the selectrons, then selectron
pair-production rates are increased.  
Of course, {\em overall} slepton
production rates go up as more slepton species are made light.
Supergravity
models predict $\msll > \mslr$; in these scenarios the $\slep_{\scsc
R}^\pm$ pair-production rate for a given $\mslr$ will be somewhat
larger than what is shown in our figures while that for $\slep_{\scsc
L}^\pm$ pair-production will be smaller by reasoning analogous to that
for the stau effect described above.
Lowering the gluino mass,
which would result in lower chargino and neutralino masses and 
potentially open some more decay channels for the Higgs bosons,
can also reduce slepton pair production rates.  However, the effect of halving
the gluino mass from its given value was found to be quite small,
mainly due to the relative insensitivity of the LSP mass to the gluino mass.
Lowering stop masses to $\sim {\cal O}$($500 \, \hbox{GeV}$), on the
other hand, increases (mainly due to the a decrease in the 
$H^0 \rightarrow h^0 h^0$ partial decay width)
cross sections for $\tb < 3$ by about 5 - 25 \%; however, 
the LEP-1 bound on $m_A$ also increases, especially for $\tb$ less than $2$, 
and rules out some of this parameter space region. 
Our choice of stau masses also affects the region excluded by LEP-1 since
$m_{\widetilde{\tau}_1} < 45\, \hbox{GeV}$ and all other slepton
masses $> \frac{1}{2}\mz$ are possible for high $\tb$.  ( The LEP-1
excluded region is taken as $\msnu < 43\, \hbox{GeV}$ and 
$\msl^{\scsc \pm} < 45 \, \hbox{GeV}$ \cite{LEP1limits}
and depicted in Fig. 1 as dotted segments
which terminate where slepton masses become unphysical.)

For $\msnu = 90\, \hbox{GeV}$, Fig. 2($a$) illustrates the
region in the $(m_A$ -- $\tb)$ plane for which the contribution from
Higgs exchange to sneutrino pair-production is comparable to 
Drell-Yan production. The solid lines show contours of the ratio $R
\equiv \sigma_{Higgs}/\sigma_{Drell-Yan}$ when $R \geq 1$, while 
dashed lines show $R < 1$. The region with relatively low values
of both $m_A$ and $\tb$ favors gluon fusion via Higgs exchange (and
also corresponds to subdominant $b \bar{b}$ annihilation).  This is
because the factor $\cos (\beta + \alpha)$ which appears in the
coupling of the heavier scalar $H^0$ to sleptons is large.  The
correspondingly small value of $\sin (\beta + \alpha)$ suppresses the coupling
of the light $h^0$ to sleptons\footnote{The $F$-term contributions to
the couplings which do not contain these factors are suppressed by
$m_{\ell}/\mz$.}. In contrast, higher values of $m_A,\tb$ suppress
the couplings of the $H^0$ and increase those of the $h^0$. The 
latter, however, mediates contributions which are always smaller than the
Drell-Yan term, essentially because the upper bound of about 
$130 \, \hbox{GeV}$ \cite{topstoploops}
on the mass $m_h$ drives the exchanged $h^0$ off-shell for all but the
lightest slepton masses.  As a result, gluon fusion becomes
subdominant for large $\tb$.  As expected, the importance of the
essentially on-shell $H^0$-exchange declines sharply when the 
$H^0 \goes t \bar{t}$ channel
opens up.  Fig. 2($b$) shows similar contours for the pair-production
of $\sler$.  The cross-section with Higgs exchanges is no longer so
strongly dominant over Drell-Yan production as in the case for
sneutrinos (see Fig. 1($b$)) and this shows up in the smaller region
in parameter space where the former is significant.  The situation is
somewhat worse for $\slel$, as shown in Fig. 2($c$).
No LEP-1 excluded region is present in Fig. 2 with our choice of 
stop masses $\simeq 1\, \hbox{TeV}$, though a large part of the region 
below $\tb \sim 2$ becomes excluded if stop masses are lowered to
$300$-$500\, \hbox{GeV}$ as discussed in the preceeding paragraph.

Figs. 3 ($a$-$c$) show contours of the Higgs-mediated cross-section
in the ($m_A,\msl$) plane for the pair production of sneutrinos,
right selectrons and left selectrons, respectively, for a fixed $\tb$.  
The region to
the left of (below) the vertical (horizontal) dash-dot line is
excluded by LEP-1 limits on $h^0$ (sleptons) 
\cite{LEP1limits}.  
The thick line on each plot corresponds to $m_H = 2\msnu$ and delineates 
the region of off-shell $H^0$-exchange where the cross-sections are seen 
to be quite small.  The contours are rather similar in each case, as
might be expected, though sneutrino pair-production has the largest
cross-sections.
Assuming (conservatively) an integrated luminosity
of $10 \, \hbox{fb}^{-1}$ at the LHC, Fig. 3($a$) shows that if 
$m_A \sim 500 \, \hbox{GeV}$, one can probe sneutrino masses as high as 
$150 \, \hbox{GeV}$ with
some 50 odd sneutrino pairs produced though kinematic cuts
would reduce the cross-section somewhat. For a lighter $m_A \simeq
300\, \hbox{GeV}$ thousands of events can be expected.

Thus, for sneutrino masses in the range $45-140\, \hbox{GeV}$ and
charged slepton masses in the range $80-140\, \hbox{GeV}$, {\it the
$H^0$-mediated gluon-fusion amplitude will be the dominant
contribution} at the LHC for low values of $\tb$ in the range $1-3$.
The Higgs-mediated cross-section for slepton masses in the ballpark
of 100 GeV can be as high as 1.5 pb, which is several orders of
magnitude larger than values quoted earlier
\cite{sleptonsignals,oldgluonfusion}.  Detection would best be
through a final state lepton pair (of the same flavor) and hard
${E\!\!\!/}_T$, since each slepton will decay into a lepton and the
LSP. This is especially so for the right slepton which does not have
a chargino decay mode.  The main backgrounds are $t \bar{t}$
production and $W$-boson pair-production, the former outstripping the
latter at the LHC. A combined requirement of hadronically quiet
events and isolated leptons with a cut of $p_{\hbox{}_T} > 20\,
\hbox{GeV}$ was shown in Ref. \cite{sleptonsignals} to essentially
eliminate the $t\bar{t}$ and $W^+W^-$ backgrounds.  The slepton
masses that make our signal most favorable are also broadly the
preferred masses in Ref. \cite{sleptonsignals}.  As our signal
cross-section can be considerably larger than theirs, most of their
arguments would go through. Moreover, since most of our sleptons come
from nearly on-shell $H^0$ bosons, each of them would tend to
have a harder $p_{\hbox{}_T}$ distribution than that predicted in
Ref. \cite{sleptonsignals} with a Jacobian peak.
Some relic features of such $p_{\hbox{}_T}$ characteristics may
survive in the dilepton signal, depending upon the kinematic cuts and
the amount of smearing due to the Hiigs boson's momentum.
Thus it might be possible to obtain a cleaner signal for slepton
pair-production than has been hitherto predicted.  We do not go into
more details of the signal analysis since we feel that this
calls for a separate investigation.

To summarize, direct electroweak pair-production of sleptons has been
shown to have potential advantages over that from cascade decays of
squarks and gluinos.  At the LHC, Higgs-exchange terms are not always
negligible as claimed in previous studies.  In fact, the $H^0$-mediated
contribution (mainly from $gg$ fusion, with a subdominant $b\bar{b}$
annihilation component) can win over the Drell-Yan cross-section for
low $\tb$ and a relatively low $m_A$.  The bulk of the former is via
a resonant on-shell $H^0$, off-shell effects being small. 
The resulting rates remain large for slepton masses beyond the potential 
reach of LEP-2.  Previous signal 
analyses should not only carry through but be strengthened by the higher 
rates now predicted. Thus, Higgs-mediated contributions may play an
important role in slepton searches at the LHC and need to be studied
carefully in any analysis of such signals for supersymmetry.

The authors are grateful to H. Baer, R. M. Godbole, D. K.
Ghosh, M. Guchait, and D. P. Roy for discussions
and correspondence, and to X. Tata for reading the manuscript and
providing useful comments.  The work of MB is partially funded by a National
Science Foundation grant (No: 9417188) while that of SR is partially
funded by a project (DO No: SERC/SY/P-08/92) of the Department of
Science and Technology, Government of India.


\newpage

\begin{center}
{\large\bf Table Caption}
\end{center}

\noindent {\it Table 1.} 
Percentages of a given slepton pair produced via the various
production modes.  Each slepton mass is fixed at $100\, \hbox{GeV}$
and $m_{\hbox{}_A}$ at $225 \, \hbox{GeV}$.  
Other values are as in Fig. 1.  A
dash signifies a percentage of under 0.05\%.  $W^+W^-$-fusion
contributions, which are expected to be quite small, are not
included.

\newpage
\begin{center}
{\large\bf Table 1}
\end{center}

\begin{table}[thb]
\vspace*{0.4cm}
\bigskip
\def\tstrut{\vrule height 14pt depth 4pt width 0pt}
\begin{tabular}{|cl|rrrrrrrrr|}
\hline \hline
Pair & Process & $\tan\beta = 1$ & $1.5$ & $1.75$ & $2$ & $2.5$ & $3$ & $5$ 
& $10$ & $15$ \\
\hline
$\tilde{\nu}\tilde{\nu}^*$ & $gg \rightarrow H^0$ &
 78.0 & 63.0 & 56.0 & 50.0 & 40.3 & 32.9 & 15.6 & 2.6 & 1.0 \\
 & $q\bar{q} \rightarrow H^0$ & 
 0.5 & 1.7 & 2.6 & 3.7 & 6.4 & 9.4 & 19.2 & 10.2 & 5.1 \\
 & $gg \rightarrow h^0$ & 
 -   & -   &  -  &  -  &  - & 0.1 & 0.1 & 0.2 & 0.3 \\
 & $q\bar{q} \rightarrow Z^{0*},{\gamma}^*$ &
 21.5 & 35.3 & 41.4 & 46.3 & 53.3 & 57.6 & 65.1 & 87.0 & 93.6 \\
\hline
$\tilde{e}_{\scriptscriptstyle R}^-\tilde{e}_{\scriptscriptstyle R}^+$ 
 & $gg \rightarrow H^0$ &
 67.1 & 48.1 & 40.6 & 34.7 & 26.3 & 20.8 & 9.8 & 1.6 & 0.6 \\
 & $q\bar{q} \rightarrow H^0$ & 
 0.5 & 1.3 & 1.9 & 2.5 & 4.1 & 5.9 & 12.0 & 6.1 & 3.0 \\
 & $gg \rightarrow h^0$ & 
 -   & -   &  -  &  -  & -  & 0.1 & 0.1 & 0.1 & 0.1 \\
 & $q\bar{q} \rightarrow Z^{0*},{\gamma}^*$ &
 34.2 & 50.6 & 57.5 & 62.8 & 69.6 & 73.2 & 78.1 & 92.2 & 96.3 \\
\hline
$\tilde{e}_{\scriptscriptstyle L}^-\tilde{e}_{\scriptscriptstyle L}^+$ 
 & $gg \rightarrow H^0$ &
 52.7 & 32.7 & 26.2 & 21.5 & 15.6 & 12.1 & 5.5 & 0.9 & 0.3 
$\!\!\!\!$ $^{\scriptscriptstyle \dagger }$ \\
 & $q\bar{q} \rightarrow H^0$ & 
 0.4 & 0.9 & 1.2 & 1.6 & 2.4 & 3.4 & 6.8 & 3.4 & 1.7
$\!\!\!\!$ $^{\scriptscriptstyle \dagger }$ \\
 & $gg \rightarrow h^0$ & 
 -   & -   &  -  &  -   & - & - & 0.1 & 0.1 & 0.1
$\!\!\!\!$ $^{\scriptscriptstyle \dagger }$ \\
 & $q\bar{q} \rightarrow Z^{0*},{\gamma}^*$ &
 46.9 & 66.4 & 72.6 & 76.9 & 82.0 & 84.5 & 87.6 & 95.6 & 97.9 
$\!\!\!\!$ $^{\scriptscriptstyle \dagger }$ \\
\hline\hline
\end{tabular}
\end{table}
\vskip -1.0cm
$^{\scriptscriptstyle \dagger} \!\!$
$m_{\tilde{\tau}_{\scriptscriptstyle 1}}= 41.8\, \hbox{GeV}$ for this case

\vskip 1.2cm

\newpage
\begin{center}
{\large\bf Figure Captions}
\end{center}

\noindent {\it Fig. 1}. Production cross-sections for 
($a$) sneutrino pairs ($\snu_e{\snu_e}^*$), ($b$) right selectron
pairs ($\widetilde{e}_{\scsc R}^+ \widetilde{e}_{\scsc R}^-$) and
($c$) left selectron pairs ($\widetilde{e}_{\scsc L}^+
\widetilde{e}_{\scsc L}^-$).  Solid lines show cross-sections for
Higgs exchanges (including $b \bar{b}$ annihilation plus gluon
fusion) for $\tb = 1,1.5,2,3$ respectively for $a,b,c$ while the
dashed line is the Drell-Yan contribution.  $m_A$ is fixed at $225\,
\hbox{GeV}$ and the values of other parameters are given in the text.

\bigskip\bigskip

\noindent {\it Fig. 2}. Contours of the ratio of the Higgs-exchange
cross-section to the Drell-Yan cross-section for ($a$) sneutrino
pairs ($\snu_e{\snu_e}^*$), ($b$) right selectron pairs
($\widetilde{e}_{\scsc R}^+ \widetilde{e}_{\scsc R}^-$) and ($c$)
left selectron pairs ($\widetilde{e}_{\scsc L}^+\widetilde{e}_{\scsc
L}^-$) in the ($m_A-\tb$)-plane.  Solid (dashed) lines show the
contours of ratio 1, 2, 3 and 5 (0.5,0.1 and 0.01) respectively.
Here $m_{\tilde{\ell}} = 90\, \hbox{GeV}$ in each case.

\bigskip\bigskip

\noindent {\it Fig. 3}. Contours in the $(m_{\tilde{\ell}} - m_A)$
plane (for $\tb = 1.5$) of the total cross-section for ($a$)
sneutrino pairs ($\snu_e{\snu_e}^*$), ($b$) right selectron pairs
($\widetilde{e}_{\scsc R}^+ \widetilde{e}_{\scsc R}^-$) and ($c$)
left selectron pairs ($\widetilde{e}_{\scsc L}^+ \widetilde{e}_{\scsc
L}^-$). The cross-section in fb is written next to each contour.  The
region to the left of (below) the vertical (horizontal) dash-dot line
is excluded by LEP-1 limits on $h^0$ (sleptons). The thick lines
correspond to $m_H = 2{\msl}$.


\newpage
                 \setcounter{page}{0}
                 \thispagestyle{empty}
                 \pagestyle{empty}

\centerline{}
\vskip 1.0cm
\begin{center}
{\large\bf Figure 1(a)}
\end{center}
\vskip -0.5cm

\centerline{}
\begin{figure}[htb]
\epsffile[100 390 432 560]{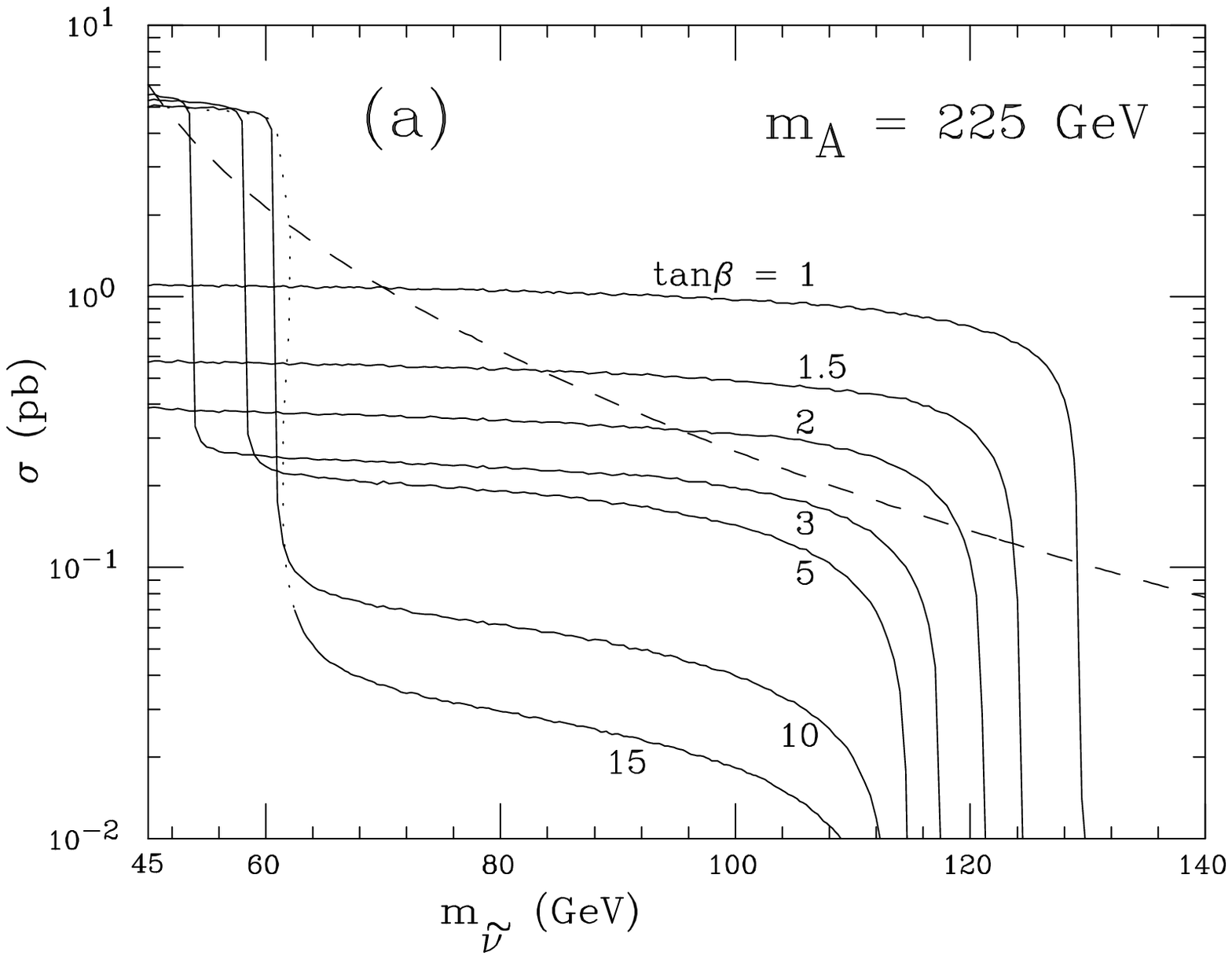}
\end{figure}

\newpage
\centerline{}
\vskip 1.0cm
\begin{center}
{\large\bf Figure 1(b)}
\end{center}
\vskip -0.5cm

\centerline{}
\begin{figure}[htb]
\epsffile[100 390 432 560]{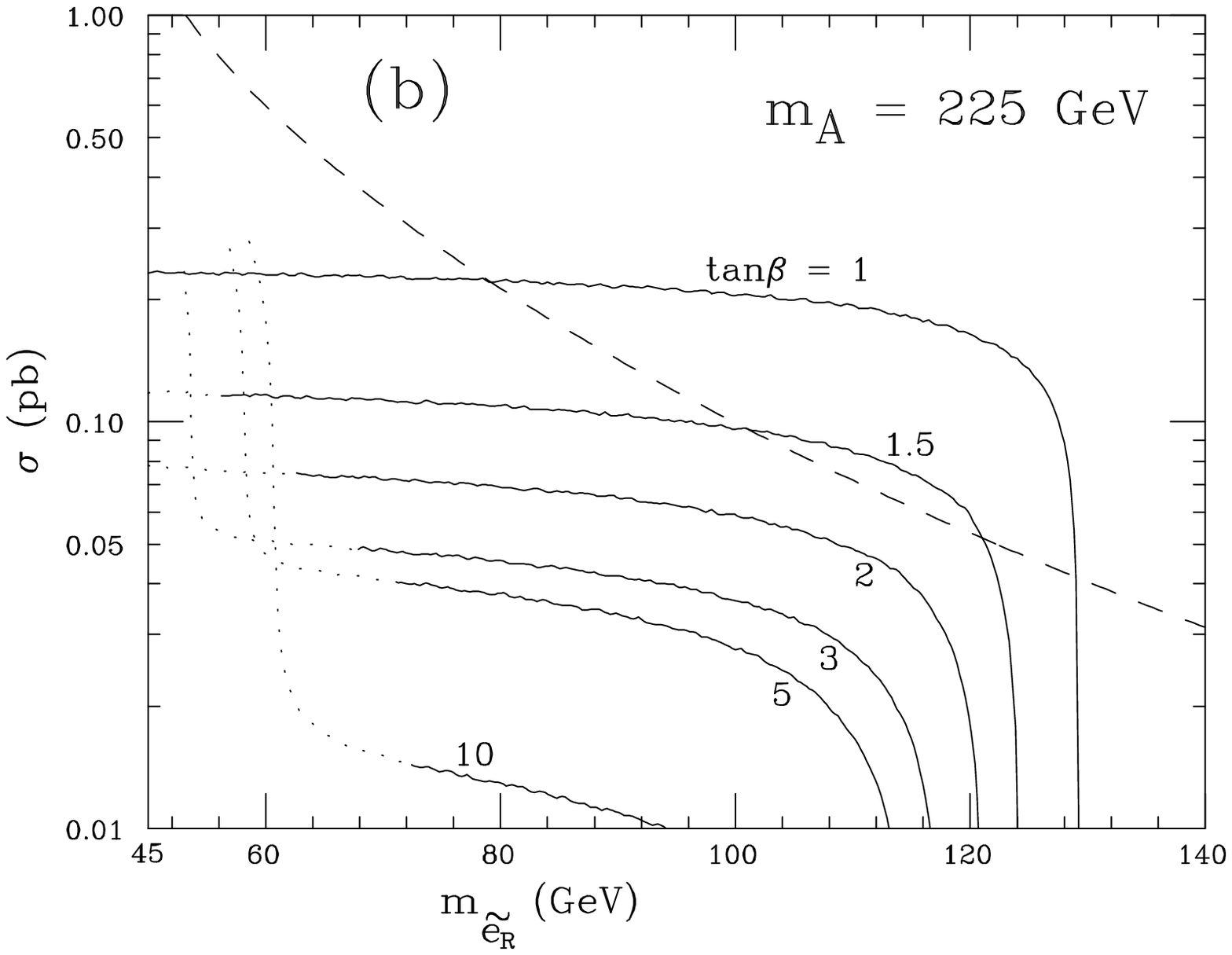}
\end{figure}

\newpage
\centerline{}
\vskip 1.0cm
\begin{center}
{\large\bf Figure 1(c)}
\end{center}
\vskip -0.5cm

\centerline{}
\begin{figure}[htb]
\epsffile[100 390 432 560]{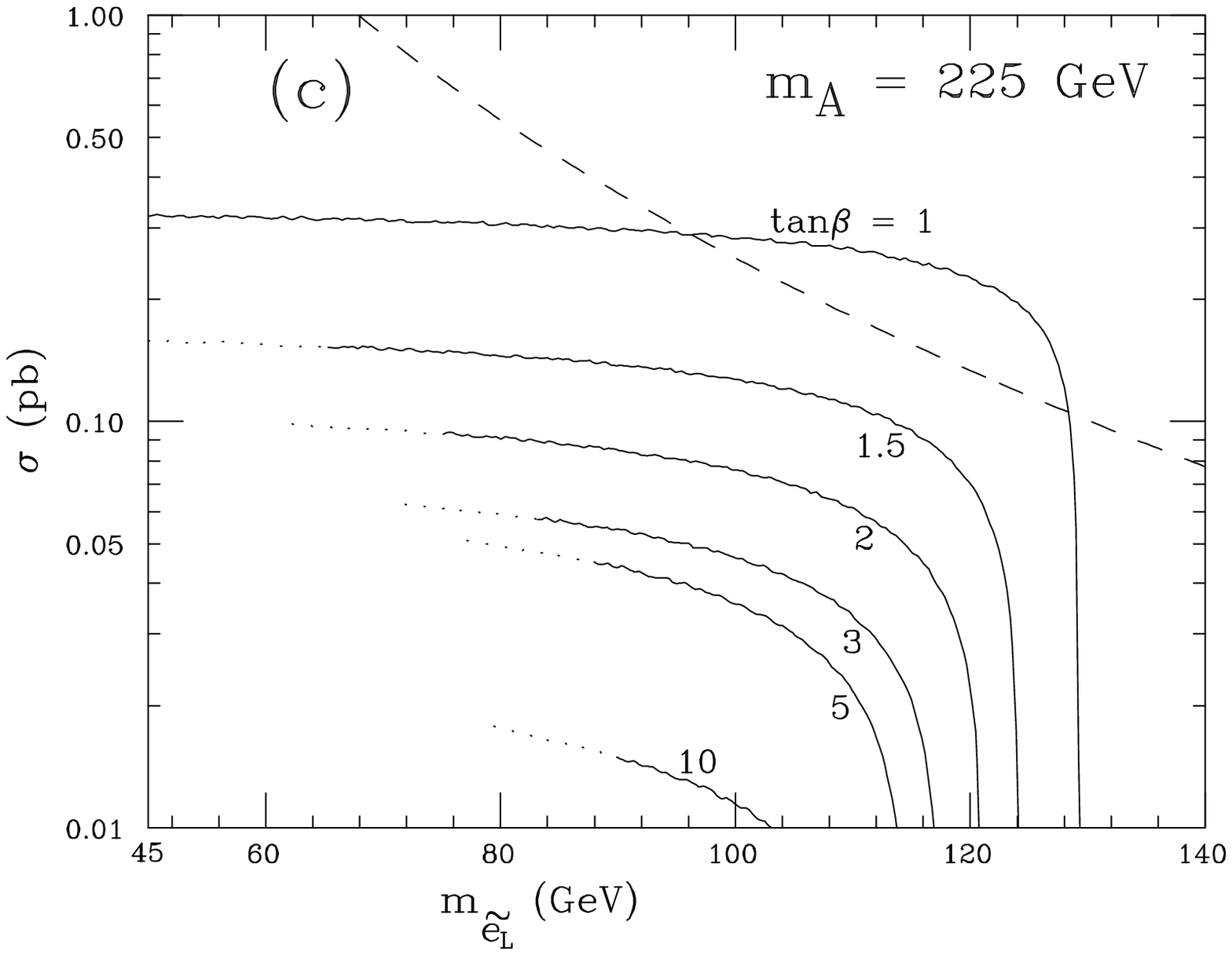}
\end{figure}

\newpage
\centerline{}
\vskip 1.0cm
\begin{center}
{\large\bf Figure 2(a)}
\end{center}
\vskip -0.5cm

\centerline{}
\begin{figure}[htb]
\epsffile[100 390 432 560]{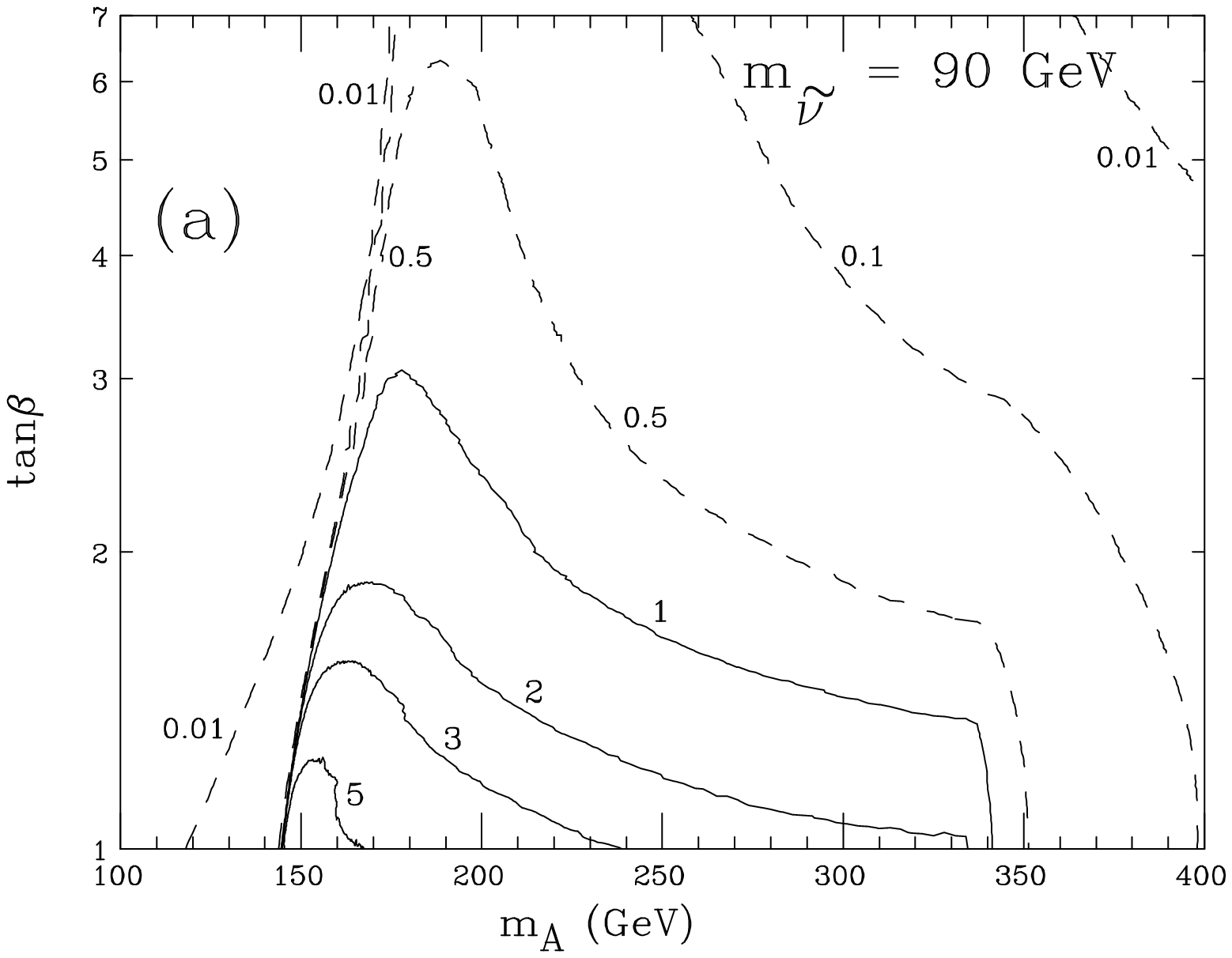}
\end{figure}

\newpage
\centerline{}
\vskip 1.0cm
\begin{center}
{\large\bf Figure 2(b)}
\end{center}
\vskip -0.5cm

\centerline{}
\begin{figure}[htb]
\epsffile[100 390 432 560]{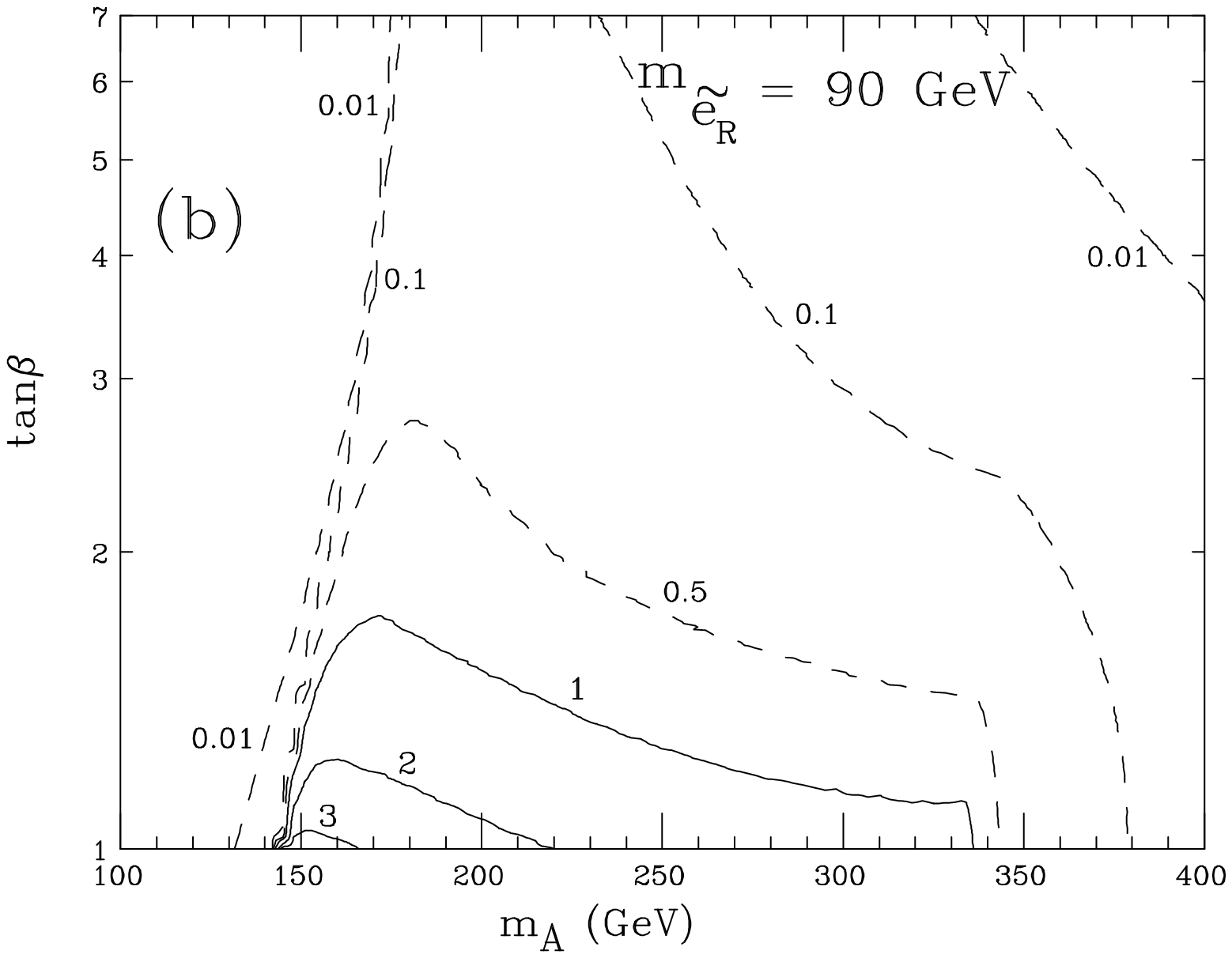}
\end{figure}

\newpage
\centerline{}
\vskip 1.0cm
\begin{center}
{\large\bf Figure 2(c)}
\end{center}
\vskip -0.5cm

\centerline{}
\begin{figure}[htb]
\epsffile[100 390 432 560]{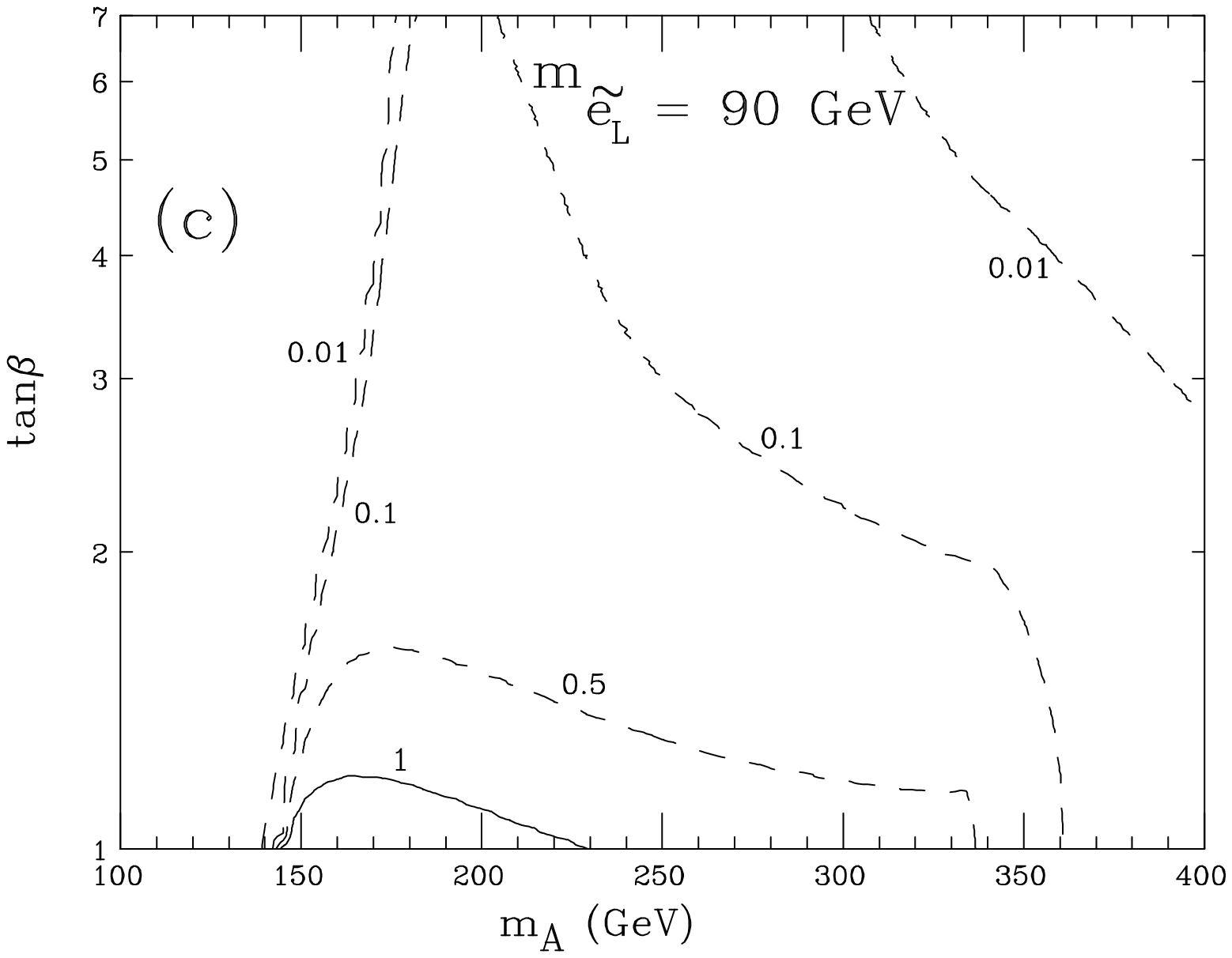}
\end{figure}

\newpage
\centerline{}
\vskip 1.0cm
\begin{center}
{\large\bf Figure 3(a)}
\end{center}
\vskip -0.5cm

\centerline{}
\begin{figure}[htb]
\epsffile[100 390 432 560]{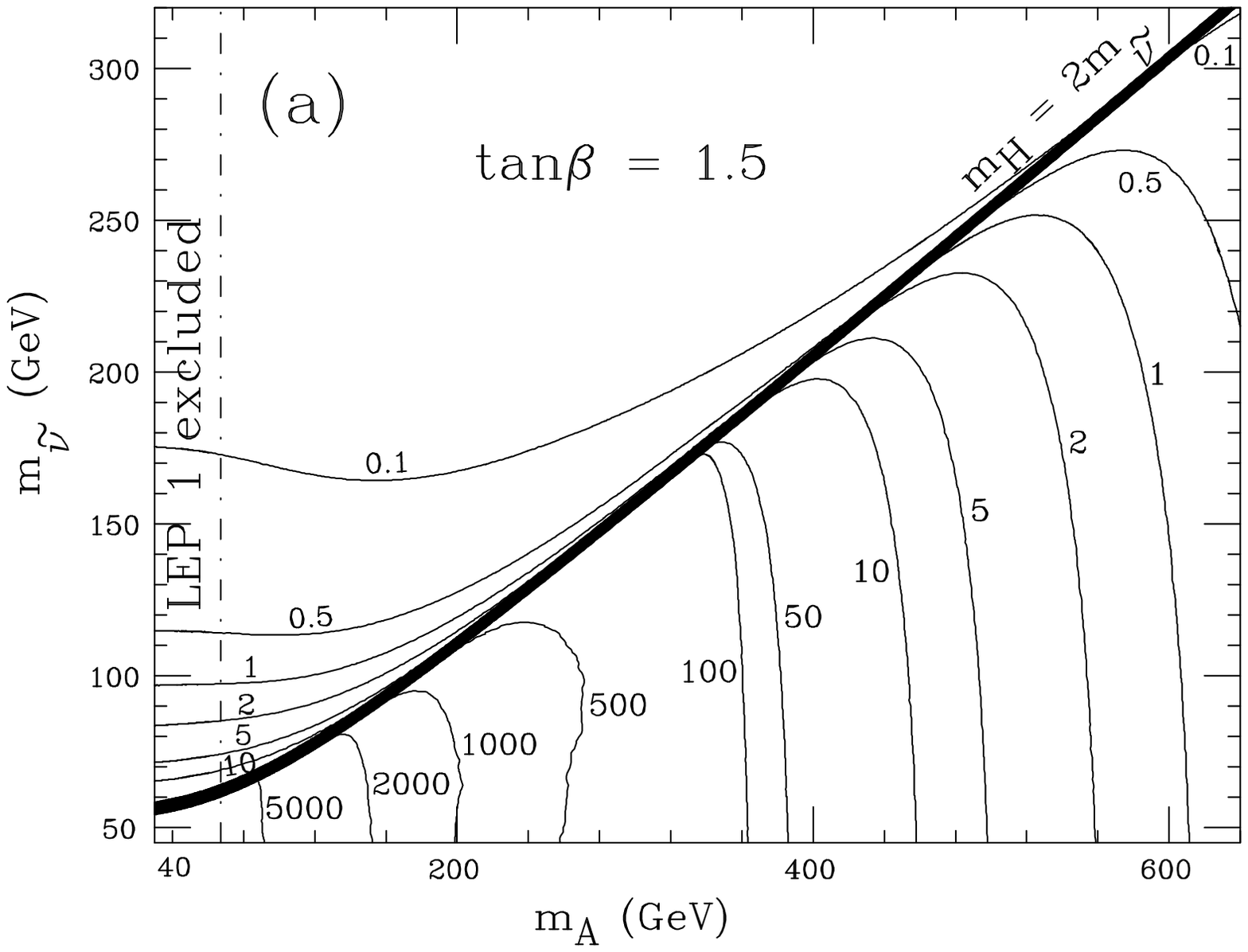}
\end{figure}

\newpage
\centerline{}
\vskip 1.0cm
\begin{center}
{\large\bf Figure 3(b)}
\end{center}
\vskip -0.5cm

\centerline{}
\begin{figure}[htb]
\epsffile[100 390 432 560]{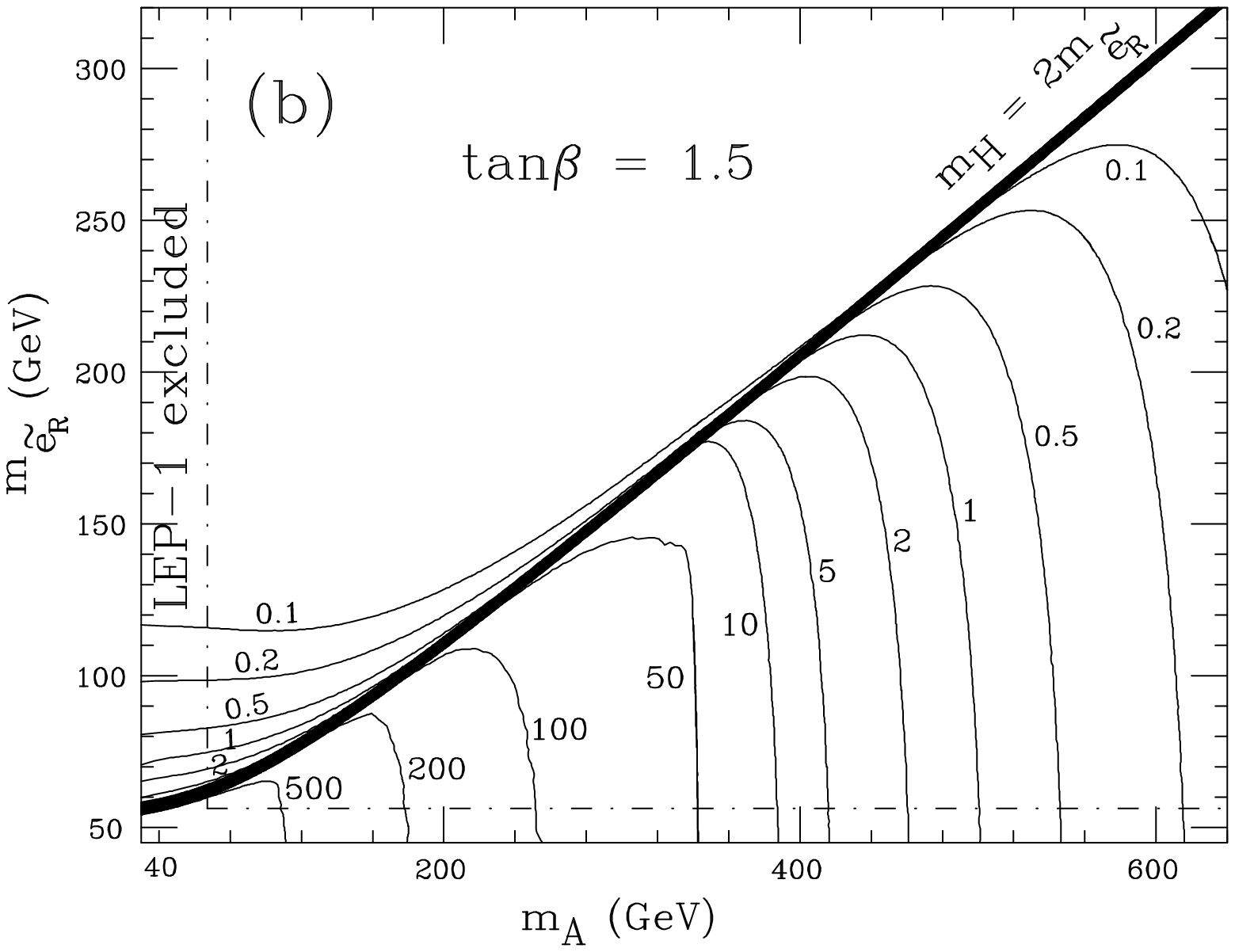}
\end{figure}

\newpage
\centerline{}
\vskip 1.0cm
\begin{center}
{\large\bf Figure 3(c)}
\end{center}
\vskip -0.5cm

\centerline{}
\begin{figure}[htb]
\epsffile[100 390 432 560]{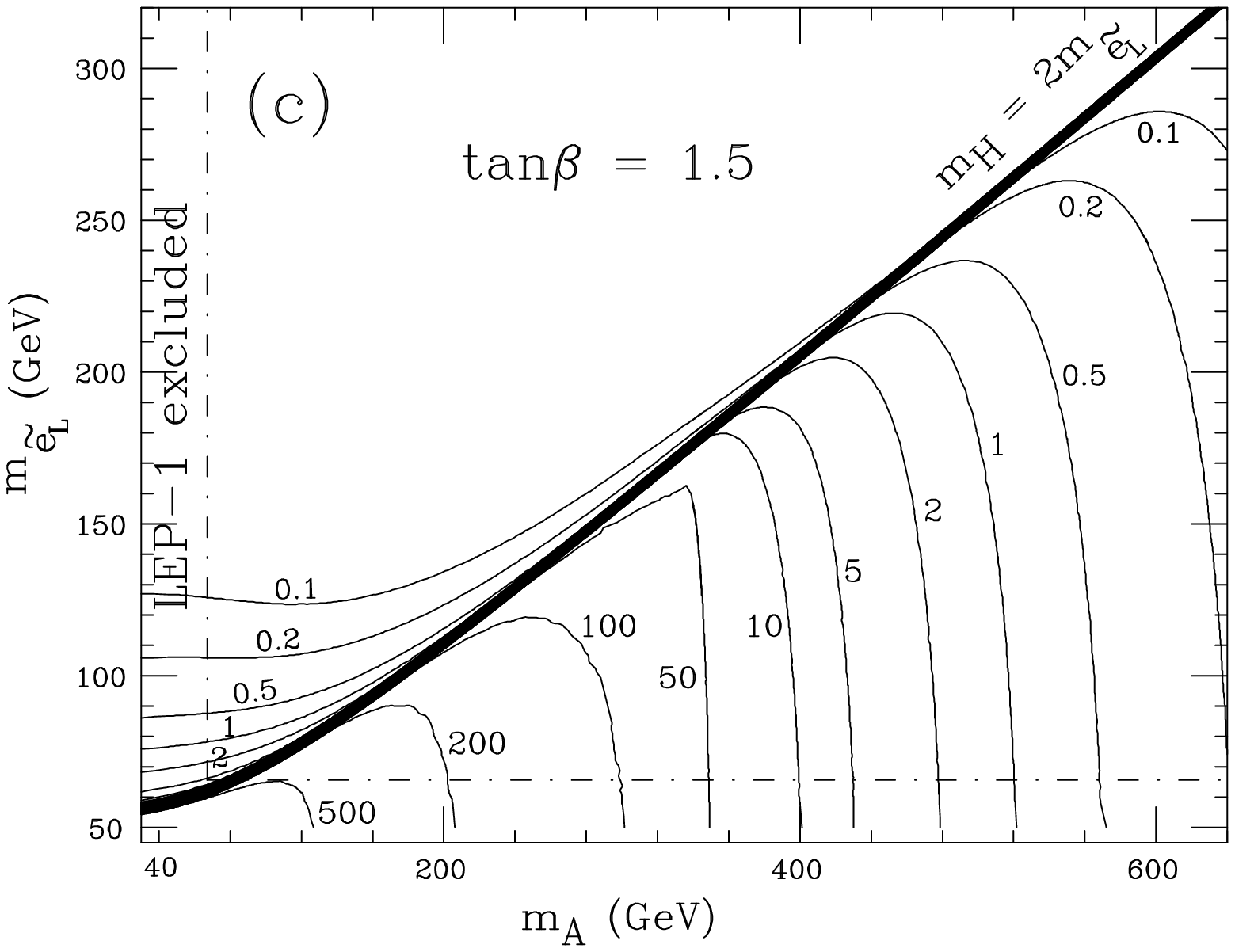}
\end{figure}


\newpage
\begin{thebibliography}{99}

\bibitem{SUSYreviews} H. E. Haber and G. L. Kane, {\it Phys. Rep.}
{\bf 117}, 1 (1985); X. Tata in {\it The Standard Model and Beyond}
ed. J.E.Kim (World Scientific, 1993); H. Baer {\it et al}., Florida State
University Report No. FSU-HEP-9504401, hep-ph/9503479; X. Tata, 
University of Hawaii Report No. UH-511-833-95, hep-ph 9510287
(1995).

\bibitem{sparticlemasses} R. Barbieri and G. F. Giudice, {\it
Nucl.Phys.} {\bf B306}, 63 (1988); G. W. Anderson and D. J. Castano,
{\it Phys.Lett.} {\bf B347}, 300 (1995).

\bibitem{EHLQ}E. Eichten, I.  Hinchliffe, K. Lane and C. Quigg, {\it
Rev.Mod.Phys.} {\bf 56}, 579 (1984); S. Dawson, E. Eichten and C. Quigg,
{\it Phys.Rev.} {\bf D31}, 1581 (1985).

\bibitem{sleptonsignals} H. Baer, C.-H. Chen, F. Paige and X. Tata,
{\it Phys.Rev.} {\bf D49}, 3283 (1994).

\bibitem{oldgluonfusion} F. del Aguila and Ll. Amettler, {\it
Phys.Lett.} {\bf B261}, 326 (1991); F. del Aguilar, Ll. Amettler,
and M. Quir\'os, Proceedings of the ECFA Workshop
on the Large Hadron Collider, eds. G. Jarlskog and D. Rein, Aachen
1990, p.663.

\bibitem{topstoploops} T. Okada, H. Yamaguchi and T. Yanagida, {\it
Prog.Theor.Phys.Lett.} {\bf 85}, 1 (1991); H. E. Haber and R.
Hempfling, {\it Phys.Rev.Lett.} {\bf 66}, 1815 (1991); J. Ellis, G.
Ridolfi and F. Zwirner, {\it Phys.Lett.} {\bf B257}, 83 (1991); P. H.
Chankowski, S. Pokorski and J. Rosiek, {\it Phys.Lett.} {\bf B274},
191 (1992).

\bibitem{cMSSM} R. Arnowitt and P. Nath in {\it Properties of SUSY
particles}: Proceedings of the VII J. A. Swieca Summer School, eds.
L. Cifarelli and V. Khoze (World Scientific, Singapore, 1993); G. L.
Kane, C. Kolda, L. Roszkowski and J. D. Wells, {\it Phys.Rev.} {\bf
D49}, 6173 (1994).

\bibitem{Guide} J. F. Gunion, H. E. Haber, G. L. Kane and S. Dawson,
{\it The Higgs Hunter's Guide} (Addison-Wesley, 1990).

\bibitem{MikeThesis} M. Bisset, {\sl Ph.D.} thesis, University of
Hawaii Report No. UH-511-813-94 (1994).

\bibitem{energydependentwidths} G. Valencia and S. Willenbrock, {\it
Phys.Rev.} {\bf D46}, 2247 (1992).

\bibitem{structurefunctions} A. D. Martin, W. J. Stirling and R. G.
Roberts, {\it Phys.Lett.} {\bf B354}, 155 (1995).

\bibitem{effectiveW} J. F. Gunion, M. H\'errero and A. Mend\'ez, {\it
Phys.  Rev.} {\bf D37}, 2533 (1988).

\bibitem{QCDcorrections} M. Spira, A. Djouadi, D. Graudenz and P. M.
Zerwas, {\it Phys.Lett.} {\bf B318}, 347 (1993).

\bibitem{DrellYanQCDcorrections} V. L. van Neerven,
{\it Int. J. Mod. Phys.} {\bf A10}, 2921 (1995).

\bibitem{radiativecorrections} D. A. Dicus and S. D. Willenbrock,
{\it Phys.Rev.} {\bf D39}, 751 (1989); H. Baer, M. Bisset, C. Kao,
and X. Tata, {\it Phys.Rev.} {\bf D50}, 316 (1994).

\bibitem{LEP1limits} L. Montanet {\it et al}.
(Review of Particle Properties), {\it Phys.Rev.} {\bf D50}, 1173 (1994);
talk by J.-F. Grivaz at Int. Europhysics Conf. on High Energy Phys.,
Brussels, July - Aug., 1995, LAL-95-83; 
see also H. Baer {\it et al}. of Ref. [1].  For methodology, see
H. Baer, M. Drees, and X. Tata, {\it Phys.Rev.} {\bf D41}, 3414 (1990);
M. Drees and X. Tata, {\it Phys.Rev.} {\bf D43}, 2971 (1991).


\end{thebibliography}
\end{document}